\documentclass[prl,amsmath,amssymb,floatfix,twocolumn,superscriptaddress]{revtex4}
\usepackage{graphicx}
\usepackage{times}

\begin{document}
\title{A topological transistor of waves via the Euler characteristic}

\author{Sophia R. Sklan}
\affiliation{Department of Mechanical Engineering, University of Colorado Boulder,
Colorado 80309 USA}
\author{Baowen Li}
\affiliation{Department of Mechanical Engineering, University of Colorado Boulder,
Colorado 80309 USA}

\begin{abstract}
Although topological materials have recently seen tremendous development, their applications have remained elusive.
Simultaneously, there exists considerable interest in pushing the limits of topological materials, including the exploration of new forms of topological protection and the establishment of topologically protected order in non-electronic systems.
Here we develop some novel forms of topological order (i.e. topological charges), primarily the Euler characteristic as well as manifold class.
We further demonstrate that these topological orders can protect bulk current transmission, even when the topologically trivial phase possesses an arbitrarily large band gap. 
Such a transition between topologically trivial, periodic dispersion and topologically non-trivial, aperiodic dispersion can be obtained by the anomalous Doppler shift of waves in a gapped periodic medium.
Since a wave’s momentum can induce an anomalous Doppler shift, we thus establish that such a transition can be used to construct a truly rigorous transistor (i.e. with switching and gain) for bosonic waves (light, sound, etc.) and that such a transistor should be experimentally realizable.
Our work suggests that additional topological charges may become relevant in moving beyond topological electronics.
\end{abstract}

\maketitle

While the field of topological materials \cite{TI} has grown tremendously, the  applications presented by these novel materials has remained limited \cite{TopoTrt4,TopoTrt5}.
This is partially due to the stringent requirements of the studied forms of topological order.
The Chern number, the principle topological charge, derives from eigenmode singularities upon an energy band.
This requires careful engineering of the band structure.
Moreover, the principle application of a non-trivial Chern number is the existence of chiral edge modes across a band gap \cite{TI2}.
While these modes' protection against backscattering is useful, the restriction to edge modes limits the currents which can travel through a topological device.

However, the Chern number is not the only topological invariant.
Other invariants exist and can impart very different topological protections to a system.
In this work, we shall examine topological connectivity instead of topologically singular points, employing the Euler characteristic (and manifold class) to develop topological devices.
The Euler characteristic is equivalent to the top Chern class on closed manifolds, but unlike that invariant it is also defined on open manifolds.

\begin{figure*}[!t]
\begin{center}
\includegraphics[scale=0.6]{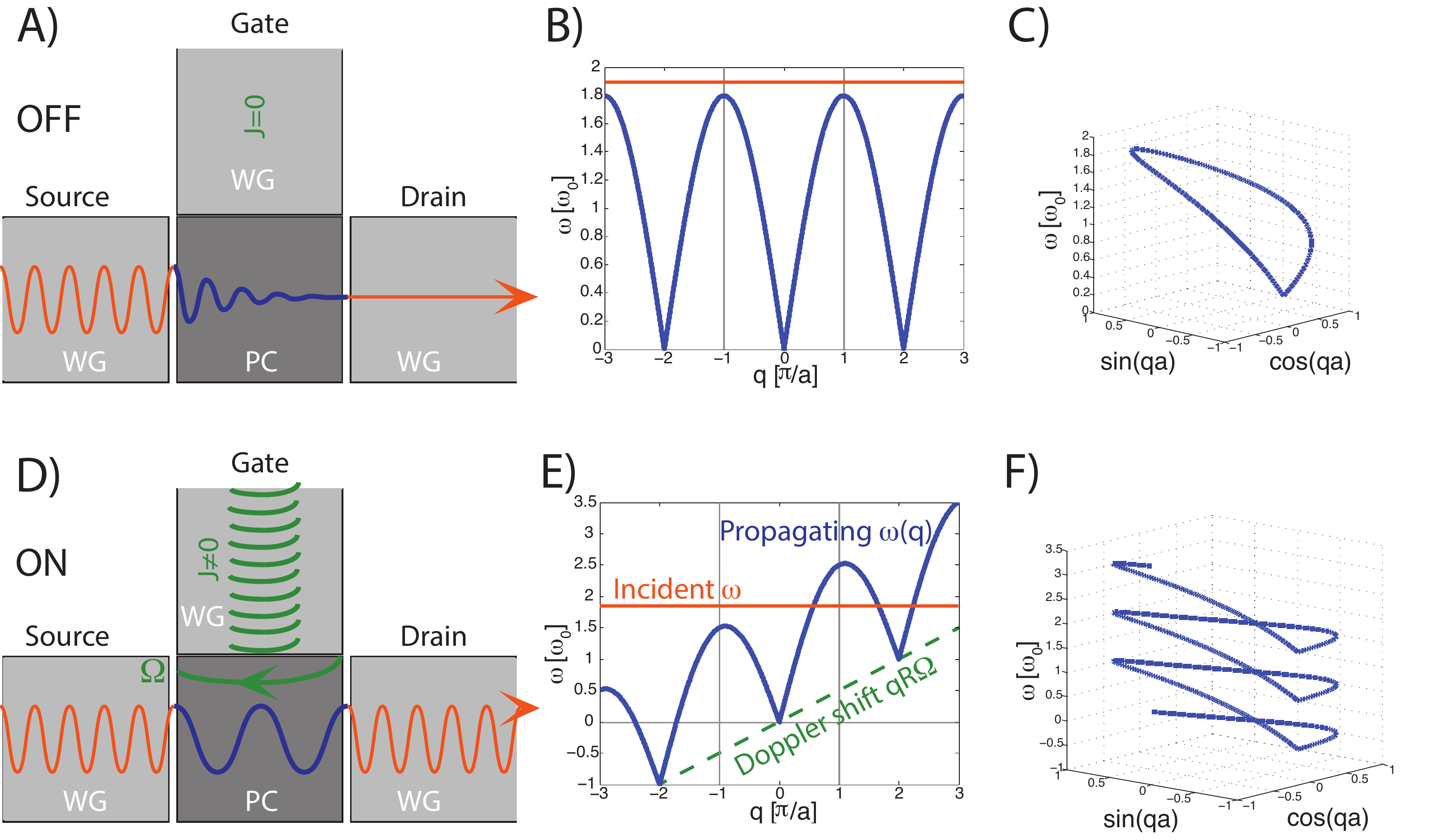}
\caption{\label{fig:model} Schematic model of the topological transistor. (A) Transistor in the ``off'' state. With no signal applied to the gate, the active site of the transistor (PC, dark grey box) is at rest relative to the source and drain (WG, light grey boxes). The wave injected into the WG (orange curve) is converted into a decaying mode in the PC (blue curve) and blocked.
(B) Dispersion relation for the PC (blue curve) in the ``off'' state, extended for several periods of the 1D BZ (black lines denote zone boundaries). Orange curve indicates the WG mode, absence of an intersection between the curves implies the nonexistence of a propagating mode.
(C) Dispersion relation of the PC in the ``off'' state, folded along the boundary of the Brillioun zone to illustrate its topological equivalence to a circle.
(D) Transistor in the ``on'' state. The signal applied to the gate (green curve) carries angular momentum $J\ne0$, inducing a rotational velocity $\Omega$ in the PC, thereby inducing DS of the wave as it enters the PC and allowing it to be transmitted.
(E) Dispersion relation for the PC in the ``on'' state, which now intersects with the WG mode due to the DS (green line).
(F) Dispersion relation of the PC in the ``on'' state, folded along the boundary of the Brillioun zone to illustrate its topological equivalence to the real line.}
\end{center}
\end{figure*}

First, recall that a Brillouin zone (BZ) is defined as a periodic space, dual to a lattice.
Periodicity exists in every direction, so an $n-$dimensional BZ is topologically equivalent to an $n-$dimensional torus, $T^n=(S^1)^{\times n}$ (see Fig. \ref{fig:model}C).
Now consider an $n-$dimensional topological product space $M^n$ where $M$ is a 1D manifold (e.g. $q_x$).
By manifold classification theory, any connected 1D manifold is equivalent to either a circle or the real line, and are distinguishable via the Euler characteristic
\begin{equation}
\chi_E=\sum_n(-1)^n b_n=\sum_n(-1)^n c_n
\end{equation}
where $b_n$ are the Betti numbers (the number of $n-$dimensional holes in the manifold) and $c_n$ are the CW complexes (the number of $n-$dimensional closed weak topology (CW) cells required to recreate the manifold).
If $M=S$ then $\chi_E=0$ as our circle has one surface ($b_0=1$) and one 1D hole ($b_1=1$) and our manifold is a compact space.
Conversely, if $M=R$, then $\chi_E=1$ as it can be reproduced with a single 1D polygon ($c_1=1$) and is a noncompact space.
Our topological product space, then, will either be a compact torus $T^n$ or a noncompact Euclidean space $R^n=(R^1)^{\times n}$ with characteristic $0^n$ or $1^n$.
Any continuous function $f:M^n\to R^1$ (e.g. a single band $\omega_a(q)$ for a toroidal BZ) defined over a compact space is necessarily bounded (there exists finite $f_{min}$ and $f_{max}$ such that, $\forall x\in M^n:f_{min}<f(x)<f_{max}$).
This is true regardless of the constraints of periodicity (although we can map a function with compact support to a periodic function over $R^n$, as in going between the reduced and extended zone schemes).
Thus any eigenmode $\omega_{a}(q),q\in T^n$, cannot exist for some range of frequencies and any finite number of bands will produce gaps.
For noncompact spaces, there is no constraint upon the boundedness of continuous functions, so the possibility of a mode existing at a given frequency is undetermined.

Now we apply a coupling to our periodic system to break periodicity along some direction.
Anticipating later results, we consider a coupling with energy proportional to $q_\Omega$, where $\Omega$ here indicates one component of $\vec{q}$ (results will hold for any unbounded energy function).
This coupling has effectively cut our manifold, i.e. along $\Omega$ the submanifold has gone from $S^1$ (Fig. \ref{fig:model}C) to $R^1$ (Fig. \ref{fig:model}E) and our BZ is $S^{n-1}\times R^1$ ($\chi_E=0^n\cdot1$), removing the boundedness constraint upon $\omega$ along this direction.
For an arbitrary periodic dispersion $\omega_{a}(q), q\in T^n$ and coupling constant $R\Omega$, we define $\omega_{max}=\omega_{a,max}+q_{\Omega}R\Omega$ and $\omega_{min}=\omega_{a,min}-q_{\Omega}R\Omega$.
Crucially, $\omega_{min}$ and $\omega_{max}$ are linear functions of $q_\Omega$ and span $\omega\in R^{1}$.
Because $\omega_{min}(q_{\Omega}R\Omega)<\omega(\omega_{a},q,\Omega)<\omega_{max}(q_{\Omega}R\Omega)$, we can use the intermediate value theorem to conclude that there must be at least one $q_{\Omega}\in R^1$ such that $\omega(\omega_a(q),q,\Omega)=\omega_D$ for arbitrary $\omega_{D}\in R^1$.
Thus, our change in the topology of the BZ allows us to ensure the existence for all $\omega$ of a surjective mapping to at least one propagating mode $\omega(q)$ in the extended BZ.
That is, it is the surjectivity of the mapping of a wave from the external environment to a propagating mode in our system that is topologically protected, rather than any individual modes.
This introduction of surjectivity in the correspondence between environment and system modes, illustrated going from Fig. \ref{fig:model}B (a non-surjective mapping) to Fig. \ref{fig:model}E (a surjective mapping) has an important corollary.
Because the propagating modes maintain amplitude whereas gapped modes decay, there will always be some minimum thickness of our system beyond which the intoduction of this coupling will increase the gap mode transmission, even if it were to increase insertion losses.

This formal result should be applicable to periodic bipartite systems (i.e. environment and system) of the form
\begin{eqnarray}
H_{sys}&=&\sum_{ai}\Pi_{ai}^2/2M_a+V(X_a)+\sum_\alpha V_{\alpha}^{cpl} \\
H_{env,\alpha}&=&\sum_{bj}P_{bj}^2/2M_b+V_\alpha(Q_b)-V_\alpha^{cpl} \\
V^{cpl}&=&\sum_{ab}V(X_a-f(t),Q_b,t)
\end{eqnarray}
where $a,b$ are site indices, $i,j$ are dimension indices, $\alpha$ environment coupling index, $\Pi=P-ne\vec{A}/c-M_a\Omega$ canonical momentum, $P$ kinematic momentum, $X$ and $Q$ are canonical positions, and $V$ is a periodic potential.
For concreteness, we consider a toy model of this form $-$ a periodic medium (PC), e.g. a photonic crystal, for a system and an environment of a series of waveguides (WG) used to inject and extract waves of a desired frequency $\omega$.
To produce a coupling of the desired form, our PC will rotate at fixed angular velocity $\Omega$.
It has been shown that moving media effect wave transport (\cite{DSPC1,DSPC2,DSPC3,DSPC4} and notably \cite{AnomDS}), the topological effect revealed by our design has not been previously examined.
Our results apply to any bosonic wave although our subsequent discussion will specialize to mechanical waves.

In this letter, we use a 1D ring of $N$ masses with mass $m$ joined by springs of stiffness $k$ (natural frequency $\omega_0^2=k/m$) and rotating with angular velocity $\Omega$ to illustrate the effect of changing topological manifold class on transport.
While the ring is 1D, the displacement ($u$) is in 3D space, so the equation of motion for the masses in the rotating frame is
\begin{equation}
m\ddot{u}_i=k(u_{i-1}+u_{i+1}-2u_i)-2m\Omega\times\dot{u}_i-m\Omega\times(\Omega\times u_i)
\end{equation}
where $i$ the site index.
The analytic solution gives
\begin{equation}
\frac{\omega^2}{\omega_0^2}=4\sin^2\frac{qa}{2}\equiv\frac{\omega^2_q}{\omega_0^2}
\end{equation}
(where $\omega$ is measured in the rotating frame, $a$ the natural separation between masses, and $2\pi/q$ the quantized wavelength) for the out-of-plane (z-polarized) mode and
\begin{equation}
\frac{\omega^2}{4\omega_0^2}=\sin^2\frac{qa}{2}+\frac{3}{8}\eta\pm\sqrt{\eta\sin^2\frac{qa}{2}+\frac{9}{64}\eta^2}
\end{equation}
($\eta=\Omega^2/\omega_0^2$) for the in-plane modes.
Since there are points in the BZ where the in-plane modes do not exist, we restrict our attention to the z-polarized mode (which exists throughout the BZ).

We model the WGs as stationary linear chains of mass $m$ and stiffness $k_e$.
Crucially, $k_e>k$ so that there exist propagating modes in the WGs which correspond to evanescent modes in the PC (i.e. above the band edge).
The WGs can be arbitrarily large, but we use $N$ masses for each chain and approximate an infinitely large chain via a damping constant $\gamma$.
These WGs are coupled to the PC by springs of stiffness $k_L=\sqrt{k_ek}$, to minimize reflection.
Because the WGs are stationary while the PC is in motion, the coupling term is
\begin{equation}
H_I^{(j)}=\frac{1}{2}\left(U_0^{(j)}-\sum_{n=pN}^{(p+1)N} u_n F\left(R\Omega t-na+R\frac{2\pi j}{\alpha}\right)\right)^2
\end{equation}
where $j$ is WG index (0 for source, and 1 and 2 for a pair of drains (see Supplement) \cite{footn,CircRect}, so $\alpha=3$), $U$ is WG displacement, the ring's radius is $2\pi R= Na$, $F$ is a Dirac delta function for analytic results and a Gaussian for numerical results, and $p$ is an integer such that $F$ remains physically relevant at arbitrary time.
Solving the full equation of motion by Fourier techniques (full derivation in Supplement) gives
\begin{gather}
(\omega^2-\omega_Q^2)u_Q(\omega)=\frac{k_{e}}{m}\frac{1}{2\pi} \sum_{q}\sum_{j=0}^{\alpha-1} e^{-iQR\frac{2\pi}{\alpha}j} \nonumber	\\ \times\left( u_q(\omega+[q-Q]R\Omega) e^{iqR\frac{2\pi}{\alpha}j}-U_q^{(j)}(\omega-QR\Omega) \right)
\end{gather}
and
\begin{equation}
U_q^{(j)}(\omega_q^{(j)})=\frac{1}{2\pi}\frac{\frac{k_{e}}{m}}{\frac{k}{m}(e^{iqa}-1)+\frac{k_{e}}{m}}u_{q}(\omega_{q}^{(j)}+qR\Omega)e^{iq\frac{2\pi}{\alpha}Rj}
\end{equation}
by which we clearly see that the rotating wave is related to the inertial mode by a Doppler shift (DS)
\begin{equation}\label{eq:Doppler}
\omega_D=qR\Omega+\omega_q
\end{equation}
where $\omega_D$ is the DSed frequency and $\omega_q$ is the dispersion relation (as anticipated in our original proof).
This is the rotational DS, which has been experimentally observed \cite{RDSExp}, and is the special case of the generalized DS
\begin{equation}\label{eq:GenDS}
\left(\omega_D-\vec{q}\cdot\vec{v}\right)^2=\omega(q)^2
\end{equation}
where $v$ is the local relative velocity \cite{AnomDS,GenDoppler}.

The intersection of each side of eq. \ref{eq:Doppler} determines the propagating PC mode.
Consider the stationary PC ($\Omega=0$), where the dispersion has the topology of a circle $S^1$ (Fig. \ref{fig:model}C).
As shown in Fig \ref{fig:model}B, when $\omega>\omega_q^{max}$, there are no intersections and therefore no propagating modes (only decaying modes).
Conversely, when $0\le\omega\le\omega_q^{max}$ there exist a pair of solutions at $\pm q(\omega)$ in the first BZ.
In fact, the BZ's periodicity implies that an infinite number of periodic solutions exist.
Contrast this with the rotating case, Fig. \ref{fig:model}E, where the dispersion has the topology of a line (Fig. \ref{fig:model}F).
The dispersion is aperiodic, so when an intersection exists it does not imply that infinite solutions also exist.
Instead, there are a finite number of solutions that bifurcate with varying $\Omega$ or $\omega$.
Furthermore, Fig. \ref{fig:model}D shows that an intersection always exists, so by our original proof there will always be at least one propagating mode excited.

Because propagating modes dissipate slower than gap modes, incoming waves in the PC's gap will have a higher amplitude for a sufficiently thick rotating PC than a stationary one.
This rotation can be driven by a second wave with non-zero angular momentum.
Using this second wave as a gate current allows us to treat our system as a transistor.
The absence (Fig. \ref{fig:model}A) or presence (Fig. \ref{fig:model}D) of the chiral gate current blocks or permits the original current's transmission.
Additionally, even a weak constant current can create a steady rotation, causing a large current at the drain (amplification).
Our system in this configuration thus meets the criteria for a transistor.
While transistor designs for photons \cite{NlinPhtTrt1,NlinPhtTrt4,EITPhtTrt1,EITPhtTrt2}, phonons or sound \cite{PhnCompRev,NlinPhnTrt2,EITPhnTrt1}, etc. exist, these approaches passed the source current through an anharmonic nonlinear medium, distorting it.  
Our design uses a parametric nonlinearity, which is effectively linear at steady-state.
Alternative transistor designs have employed transduction \cite{PhtSwitch2, MAPhnTrt,Photoisomer} produce switching but not gain. 
Gain there is limited by conversion losses, which are absent in our torque method.
Our topologically protected transistor thus has several advantages compared to previous designs.

\begin{figure}
\begin{center}
\includegraphics[scale=0.45]{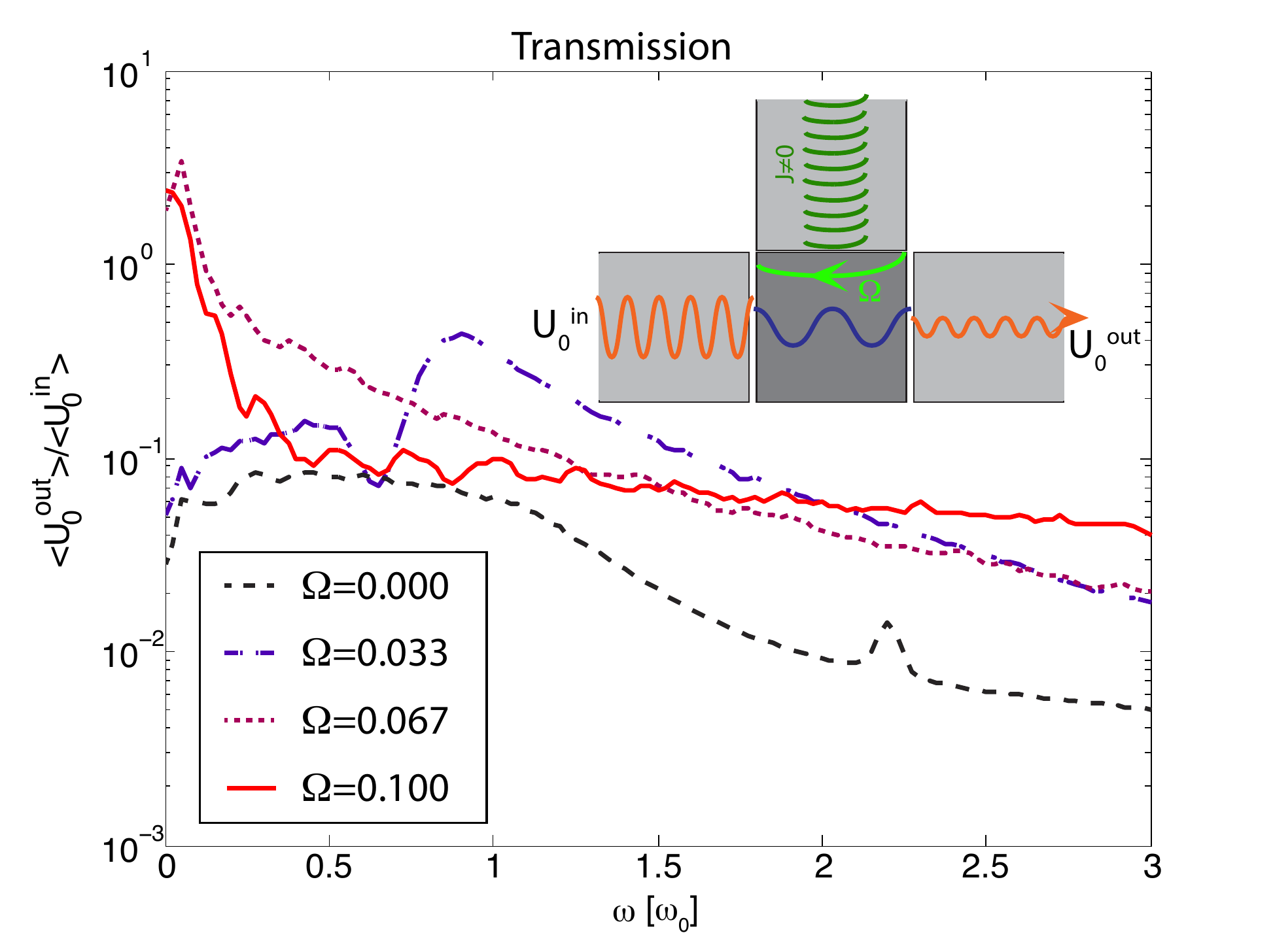} 
\caption{\label{fig:TvI} Semi-log plot of transmission (outgoing amplitude normalized by excited amplitude) along a single drain as a function of driving frequency $\omega$. Color denotes rotational velocity $\Omega$, varying between 0 and $0.1\omega_0$. Comparing the stationary case (dark blue curve) with the rotating case (other curves) reveals switching.}
\end{center}

\end{figure}

\begin{figure}
\begin{center}
\includegraphics[scale=0.45]{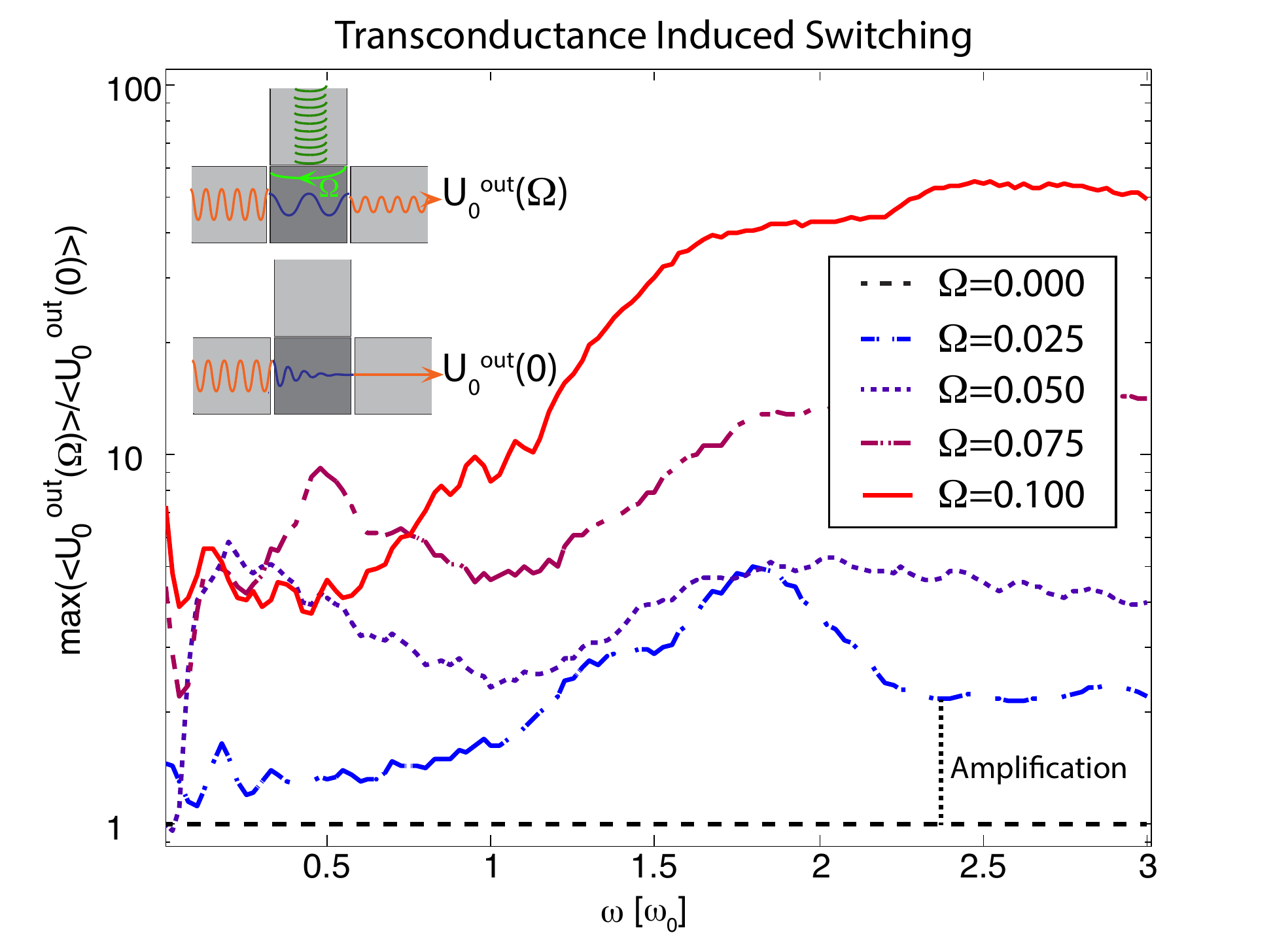} 
\caption{\label{fig:TvT} Semi-log plot of switching (outgoing amplitude normalized by stationary outgoing amplitude) along either drain as a function of driving frequency $\omega$. Color denotes rotational velocity $\Omega$, varying 0 and $0.1\omega_0$. The height of the rotating case (other curves) above the stationary case (black horizontal line) quantifies the amplification at a given $\Omega$ and $\omega$ (black vertical line). The maximum switching for the drains is used because the rectification alternates direction as a function of $\omega$ whereas switching refers to the dominant change in single channel transmission.}
\end{center}
\end{figure}

We now examine the transmitted signal.
Absolute transmission drops-off with increasing $\omega$ (see Supplement), which initially seems detrimental to our transistor, but is to be expected for constant force (rather than constant amplitude) driving.
Fig. \ref{fig:TvI} is a semi-log plot of amplitude at one drain (variation in transmission with choice of drain is discussed in the Supplement) compared to the source amplitude as a function of driving frequency for various rotational velocities $\Omega$.
Although this transmission is lossy (something that may be ameliorated with further optimization), rotating state's transmission almost always exceeds the stationary state's transmission due to the topologically protected propagation.
Normalizing the transmission $T(\omega,\Omega)$ at a given driving frequency with the stationary $T(\omega,0)$, such as is shown in Fig. \ref{fig:TvT} reveals the improved transmission as a result of the PC's rotation.
This ratio can be quite large, reaching a factor of 60 improvement in our simulations.
In principle it can be made arbitrarily large by increasing the PC size, as it is dominated by the stationary case's exponential decay.
This improvement as a result of the applied rotation implies switching $-$ shifting between the stationary and rotating states results in a large change in the transmission.


Now we consider the operation of the gate.
Because the change between stationary and rotating cases changes the band structure topology, it will exist for any non-trivial $\Omega$.
Thus, if a weak wave at the top of the ring (the gate, see Fig. \ref{fig:model}D) can induce any rotation via some angular momentum \cite{PhnAM}, the much larger signal that will be transmitted will constitute gain.
Angular momentum can be carried through the circular polarization of transverse modes for spin angular momentum \cite{PhnAM,OAM,OAM2} or the spatio-temporal patterning of the phase for orbital angular momentum \cite{OAM,OAM2,OAM3,OAM4} 
and induce a controlled rotation of an object \cite{PhnAM,OAM,OAM2,OAM3,OAM4}. 

Combining these elements, we have both switching and gain and thus a wave transistor.
In general, this approach only requires a small set of criteria.
First, the WG supports modes within the stationary PC's band gap (i.e. $k_e>k$ in our model).
Second, the PC's dispersion exists throughout the BZ (large $N$ in our model).
And third, the rotation must be induced by the same type of wave as our source (the assumption of a circularly polarized or phase-masked gate signal).

With these criteria in mind, we now consider experimentally realizing such a transistor.
For brevity, we restrict attention to optical or acoustic waves, although generalizations are straightforward.
Optical waveguides are commonplace, and creating photonic crystals with band gaps is a solved problem \cite{PhtCrys4,PhtCrys5}. 
Optically induced rotation can be due to optical torque \cite{OAM,OAM2} on mechanically stabilized (although such stabilization can entail losses or multiple drains) or optically levitated objects (accessible for sub-kg systems and length scales down to the diffraction limit) \cite{OAM2,PhtTrap,PhtTrap3}. 
While optical levitation has not been applied to a photonic crystal, several promising experiments \cite{PhtTrapArr,PhtTrapArr2} on the rotation and translation of arrays of trapped particles suggest that this is feasible.

For acoustics, waveguides are similarly feasible \cite{PhnWG,PhnWG2}, albeit less often studied.
Furthermore, band gaps have been created through phononic or sonic crystals  \cite{PhnWG2,PhnCrys3}, but remain a greater technical challenge than in optics. 
Acoustic levitation and angular momentum transfer have also been developed and are perhaps more easily employed than in optics \cite{OAM3,OAM4}, including the levitation of kg mass objects \cite{PhnTrap}. 
A combined levitation and rotation of a sonic crystal or an array of objects has also been less extensively studied than the optical case, but some preliminary works exist \cite{PhnTrapArr,PhnTrapArr2}.

Additionally, while not technically a transistor, the gate and source signals could be different types of waves.
In particular, one could be photonic and the other could be phononic.
This would take advantage of the easier fabrication of optical waveguides and photonic crystals and the wider applicability of acoustic levitation and rotation.
Moreover, such a hybrid system entails a parametric photon-phonon coupling, adding another tool to the optomechanics toolbox.

Breaking the dispersion's periodicity is a topological transition that is outside the standard topological materials framework \cite{TI}. 
This includes other realizations of a topological transition-induced transistor \cite{TopoTrt4,TopoTrt5,TopoTrt,TopoTrt2,TopoTrt3,TopoTrt6,TopoTrt9,TopoTrt0} which rely upon introduction of topological protected chiral edge modes in a bulk insulator where the valence and conduction bands possess non-trivial Chern number. 
Changing the manifold class of the band structure (e.g. band structure closure) is distinct from changing its Chern class. 
Because the Euler characteristic equals the top Chern class only for a closed manifold, the transition to an open manifold with new Euler characteristic goes beyond the Chern class topological material framework.
As such, the presence of the transistor effect in this regime strongly suggests that additional topological charges may become relevant and useful in moving beyond topological electron states.


\begin{thebibliography}{30}

\bibitem{TI}M.Z. Hasan and C.L. Kane, \textit{Rev. Mod. Phys.} \textbf{82}, 3045 (2010).

\bibitem{TopoTrt4}M. Ezawa, \textit{Appl. Phys. Lett.} \textbf{102}, 172103 (2013).
\bibitem{TopoTrt5}J. Liu, T.H. Hsieh, P. Wei, W. Duan, J. Moodera, and L. Fu, \textit{Nat. Mater.} \textbf{13}, 178 (2014).

\bibitem{TI2} L. Lu, J. D. Joannopoulos, and M. Solja\v{c}i\'{c}, \textit{Nat. Photon.} \textbf{8}, 821 (2014). 

\bibitem{DSPC1}K. Karrai, I. Favero, and C. Metzger, \textit{Phys. Rev. Lett.} \textbf{100}, 240801 (2008). 
\bibitem{DSPC2}E.J. Reed, M. Solja\v{c}i\'{c}, and J.D. Joannopoulos,  \textit{Phys. Rev. Lett.} \textbf{91}, 133901 (2003). 
\bibitem{DSPC3}J. Chen, Y. Wang, B. Jia, T. Geng, X. Li, L. Feng, W. Qian,
B. Liang, X. Zhang, M. Gu, and S. Zhuang, \textit{Nat. Photon.} \textbf{5}, 239–245 (2011). 
\bibitem{DSPC4}D.W. Wang, H.T. Zhou, M.J. Guo, J.X. Zhang, J. Evers, and S.Y. Zhu, \textit{Phys. Rev. Lett.} \textbf{110}, 093901 (2013). 
\bibitem{AnomDS}X. Hu, Z. Hang, J. Li, J. Zi, and C.T. Chan, \textit{Phys. Rev. E} \textbf{73}, 015602(R) (2006). 

\bibitem{footn} In practice it may be desirable to have $\alpha$ multiple drains, e.g. to stabilize and confine the rotating PC or to create a rectification effect \cite{CircRect} as analyzed in the Supplement.
\bibitem{CircRect}R. Fleury, D.L. Sounas, C.F. Sieck, M.R. Haberman, and A. Al\`{u},  \textit{Science} \textbf{343}, 516 (2014).
\bibitem{RDSExp}K.D. Skeldon, C. Wilson, M. Edgar, and M.J. Padgett, \textit{New J. Phys.} \textbf{10}, 013018 (2008).
\bibitem{GenDoppler}M.W. Dingemans, \textit{Water wave propagation over uneven bottoms. Advanced Series on Ocean Engineering 13}, (World Scientific, Singapore, 1997).

\bibitem{NlinPhtTrt1}K. Jain and G.W. Pratt Jr, \textit{Appl. Phys. Lett.} \textbf{28}, 719 (1976).
\bibitem{NlinPhtTrt4}M.F. Yanik, S. Fan, M. Solja\v{c}i\'{c}, and J.D. Joannopoulos, \textit{Opt. Lett.} \textbf{28}, 2506 (2003).

\bibitem{EITPhtTrt1}J. Hwang, M. Pototschnig, R. Lettow, G. Zumofen, A. Renn, S. G\"{o}tzinger, and V. Sandoghdar, \textit{Nature} \textbf{460}, 76 (2009).
\bibitem{EITPhtTrt2}W. Chen, K.M. Beck, R. B\"{u}cker, M. Gullans, M.D. Lukin, H. Tanji-Suzuki, and V. Vuleti\'{c}, \textit{Science} \textbf{341}, 768 (2013).

\bibitem{PhnCompRev}S.R. Sklan, \textit{AIP Advances} \textbf{5}, 053302 (2015).
\bibitem{NlinPhnTrt2}F. Li, P. Anzel, J. Yang, P.G. Kevrekidis, and C. Daraio, \textit{Nat. Commun.} \textbf{5}, 5311 (2014).

\bibitem{EITPhnTrt1}D. Hatanaka, I. Mahboob, K. Onomitsu, and H. Yamaguchi, \textit{Appl. Phys. Lett.} \textbf{102}, 213102 (2013).

\bibitem{PhtSwitch2}D.E. Chang, A.S. S\o rensen, E.A. Demler, and M.D. Lukin, \textit{Nat. Phys.} \textbf{3}, 807 (2007).

\bibitem{MAPhnTrt}S.R. Sklan and J.C. Grossman, \textit{New J. Phys.} \textbf{16}, 053029 (2014).
\bibitem{Photoisomer}S.R. Sklan and J.C. Grossman, \textit{Phys. Rev. B} \textbf{92}, 165107 (2015).

\bibitem{PhnAM}A.G. McLellan, \textit{J. Phys. C Solid State} \textbf{21}, 1177 (1988).
\bibitem{OAM}A.M Yao and M.J. Padgett, \textit{Adv. Opt. Photonics} \textbf{3}, 161 (2011).
\bibitem{OAM2}H. He, M.E.J. Friese, N.R. Heckenberg, and H. Rubinsztein-Dunlop, \textit{Phys. Rev. Lett.} \textbf{75}, 826 (1995).
\bibitem{OAM3}A. Marzo, S.A. Seah, B.W. Drinkwater, D.R. Sahoo, B. Long, and S. Subramanian, \textit{Nat. Commun.} \textbf{6}, 8661 (2015).
\bibitem{OAM4}R. Wunenburger, J.I.V. Lozano, and E. Brasselet, \textit{New J. Phys.} \textbf{17}, 103022 (2015).

\bibitem{PhtCrys4}A. Blanco, E. Chomski, S. Grabtchak, M. Ibisate, S. John, S. W. Leonard, C. Lopez, F. Meseguer, H. Miguez, J.P. Mondia, and G.A. Ozin, \textit{Nature} \textbf{405}, 437 (2000).
\bibitem{PhtCrys5}P. Russell, \textit{Science} \textbf{299}, 358 (2003).

\bibitem{PhtTrap}S. Chang, and S.S. Lee, \textit{J. Opt. Soc. Am. B} \textbf{2}, 1853 (1985).
\bibitem{PhtTrap3}O.M. Marag\`{o}, P.H. Jones, P.G. Gucciardi, G. Volpe, and A.C. Ferrari, \textit{Nature Nanotechnol.} \textbf{8}, 807 (2013).
\bibitem{PhtTrapArr}S. Kuhn, P. Asenbaum, A. Kosloff, M. Sclafani, B.A. Stickler, S. Nimmrichter, K. Hornberger, O. Cheshnovsky, F. Patolsky, and M. Arndt, \textit{Nano Lett.} \textbf{15}, 5604 (2015).
\bibitem{PhtTrapArr2}J. Moore, L.L. Martin, S. Maayani, K.H. Kim, H. Chandrahalim, M. Eichenfield, I.R. Martin, and T. Carmon, \textit{Opt. Express} \textbf{24}, 2850 (2016).

\bibitem{PhnWG}A. Khelif, A. Choujaa, S. Benchabane, B. Djafari-Rouhani, V. Laude, \textit{Appl. Phys. Lett.} \textbf{84}, 4400 (2004).
\bibitem{PhnWG2}T. Miyashita, \textit{Meas. Sci. Technol.} \textbf{16}, R47 (2005).

\bibitem{PhnCrys3}J.O. Vasseur, P.A. Deymier, B. Chenni, B. Djafari-Rouhani, L. Dobrzynski, and D. Prevost, \textit{Phys. Rev. Lett.} \textbf{86}, 3012 (2001).

\bibitem{PhnTrap}S. Ueha, Y. Hashimoto, and Y. Koike, \textit{Ultrasonics} \textbf{38}, 26 (2000).
\bibitem{PhnTrapArr}Y. Tian and R.E. Apfel, \textit{J. Aerosol Sci.} \textbf{27}, 721 (1996).
\bibitem{PhnTrapArr2}K. Melde, A.G. Mark, T. Qiu, and P. Fischer, \textit{Nature}, \textbf{537}, 518 (2016).

\bibitem{TopoTrt}J. Maciejko, E.A. Kim, X.L. Qi, \textit{Phys. Rev. B} \textbf{82}, 195409 (2010).
\bibitem{TopoTrt2}V. Krueckl and K. Richter, \textit{Phys. Rev. Lett.} \textbf{107}, 086803 (2011).
\bibitem{TopoTrt3}L.A. Wray, \textit{Nat. Phys.} \textbf{8}, 705 (2012).
\bibitem{TopoTrt6}M. Ezawa, \textit{New J. Phys.} \textbf{16}(6): 065015 (2014).
\bibitem{TopoTrt9}Q. Liu, X. Zhang, L.B. Abdalla, A. Fazzio, and A. Zunger, \textit{Nano Lett.} \textbf{15}, 1222 (2015).
\bibitem{TopoTrt0}K. Joulain, J. Drevillon, Y. Ezzahri, and J. Ordonez-Miranda, \textit{Phys. Rev. Lett.} \textbf{116}, 200601 (2016).

\end{thebibliography}
\end{document}